\documentclass[a4paper,12pt]{article} 
\usepackage[T1]{fontenc}
\usepackage[utf8]{inputenc} 
\usepackage{amsmath} 
\usepackage{amsfonts}
\usepackage{amssymb} 
\usepackage[english]{babel}
\usepackage{graphicx,color}
\title{Domain wall interactions due to vacuum Dirac field
fluctuations in $2+1$ dimensions} 
\author{C.~D.~Fosco\\
	and\\
	F. D. Mazzitelli\\ \\
{\normalsize\it Centro At\'omico Bariloche and Instituto Balseiro}\\
{\normalsize\it Comisi\'on Nacional de Energ\'\i a At\'omica}
\\
{\normalsize\it 8400 Bariloche, Argentina}}
\begin{document}
\date{}
\maketitle
\begin{abstract}
\noindent 
We evaluate quantum effects due to a $2$-component Dirac field in $2+1$
space-time dimensions, coupled to  domain-wall like defects with a smooth
shape.  We show that those effects induce non trivial contributions to the
(shape-dependent) energy of the domain walls.  
For a single defect, we study  the divergences in the
corresponding self-energy, and also consider the role of the massless zero
mode, corresponding to the Callan-Harvey mechanism, 
by coupling the Dirac field to an external gauge field.
For two defects, we show that the Dirac field induces a non trivial,
Casimir-like effect between them, and provide an exact expression for that
interaction in the case of two straight-line parallel defects.  As is the
case for the Casimir {\it interaction} energy, the result is finite and
unambiguous. 
\end{abstract}
\section{Introduction}\label{sec:intro}
The study of effects due to fermionic fields in the background of defects,
has been a subject of general interest in rather different areas, from
the behavior of textures in superfluid phases of ${\rm He}_3$ \cite{Ho:1984zz} to
cosmic strings \cite{Witten:1984eb}.  
A representative of those phenomena is the Callan-Harvey
effect \cite{Callan:1984sa}, where a Fermi field in an odd-dimensional
spacetime couples to a defect, the latter corresponding to a mass term which
changes sign along a domain wall.  Under these circumstances, the Dirac
field spawns a localized zero mode which may be capable of carrying
currents when coupled to a gauge field.

A similar phenomenon happens also in  a non-relativistic two-dimensional
electron gas in the presence of a magnetic field in the regime of the
quantum Hall effect, where those zero modes become the so called chiral
edge states \cite{Wen:1990se}. 

In this paper we evaluate yet another effect due to the interplay between the
fermionic field and domain walls, this time involving not just the presence
of the zero mode, but also the quantum fluctuations of the fermionic field
on top of the domain walls. This effect amounts to the emergence of non trivial
contributions to the vacuum energy as a function of the wall(s) geometry.
One should indeed expect this kind of effect, since the existence of the
walls produces a geometry dependent distortion of the vacuum fluctuation, a
fertile set up for the induction of Casimir-like effects. Indeed, the quantum vacuum 
interaction between kinks of the Sine-Gordon equation
has been evaluated  in Ref.\cite{BordagSG} using the TGTG-formula \cite{Kenneth}.
This force is attractive, and it should be considered as a small quantum
correction to the well known repulsive classical force between kinks
\cite{Manton}.  

Having in mind  its potential application to graphene~\cite{Cortijo:2012},
where some effective continuum models correspond to Dirac fields in $2+1$
dimensions coupled to space-dependent masses~\cite{Ebert:2014}, we evaluate
here some effects due to the vacuum (i.e., zero temperature) quantum
fluctuations of a Dirac field in the presence of domain walls \cite{Semenoff}.
Since there is no reason a priori to assume that just the zero modes are
relevant to this effect, we will include all the modes. 

Effective continuum models for graphene using Dirac fields generally
involve not just a single $2$-component field, but an even number of them,
what amounts to putting the Dirac field in a reducible representation of
Poincar\'e's group in $2+1$ dimensions. Nevertheless, since those models
can be constructed in terms of decoupled $2$-component fields in a rather
straightforward way, we will consider just the latter, i.e., spinors in an
irreducible representation. Besides, the $2$ component case is relevant to
other applications in condensed matter physics, like the above mentioned
quantum Hall effect case.  

The domain wall energy is usually described by means of an effective
Landau-Ginzburg like functional of its shape (assumed to be smooth). That
functional could, at least in principle, be obtained from a detailed
microscopic model for the system. In this context,  contributions depending
on, for example, the crystal structure, should be quite relevant.  We
focus, in this work, on contributions to the domain wall energy, which
should appear on top of the ones coming from the lattice structure.
Besides, we also obtain a result corresponding to two domain walls, whereby
the Dirac field is shown to induce an attractive, Casimir-like force.  This
force does not have, to the best of our knowledge, an analogue within the
context of the phenomenological model, since it is not a contact
interaction which could be incorporated by means of a local term into the
energy.

For static-wall configurations, we shall see that the effective action
predicts the existence of an effective interaction between domain-wall like
defects in the mass (the `pseudo-gap'). We also study the dependence of
that interaction as a function of the geometry of the defects, at least
under some simplifying assumptions and for particular cases.

This paper is organized as follows: in Sect.~\ref{sec:thesys} we describe
the kind of model that we consider in this article, introduce our notation
and conventions, and define its effective action. As a warming up exercise
we consider, in Sect.~\ref{sec:sw}, the case of a single domain wall,
namely, of a static mass that changes sign along a single spatial curve,
having constant values (and the same module) everywhere else.  The
corresponding self-energy is divergent, even the contributions
corresponding to small deformations of a rectilinear wall.
We also verify, by coupling the field to an external gauge field, that the massless fermion mode is properly taken into account in our approach.

Then in Sect.~\ref{sec:dw} we deal with the situation of a static mass that
changes sign on two spatial curves, also having the same constant value
elsewhere. We compute exactly the interaction energy for the particular case of two straight lines, showing that the force between defects is always attractive.
A perturbative expansion to treat cases when the shape of one of the rectilinear walls is slightly perturbed can be implemented. An Appendix contains the details corresponding to the first and second order terms.

Sect.~\ref{sec:conc} contains our conclusions.

\section{The system}\label{sec:thesys}
The system that we consider in this work is defined in terms of an
Euclidean action  ${\mathcal S}$, given by:
\begin{equation}\label{eq:defsf}
{\mathcal S}({\bar\psi},\psi,M) \;=\; \int d^3x \,\bar\psi(x) \, {\mathcal
D} \, \psi(x) \;,
\end{equation}
with ${\mathcal D} \equiv \not \!  \partial \,+\, M({\mathbf x})$, 
for a Dirac field $\psi(x), \, \bar\psi(x)$ in the presence of a space
dependent mass $M({\mathbf x})$, in $2+1$ dimensions. We have adopted the
convention that $x$ denotes the three Euclidean spacetime coordinates,
$x_0,\, x_1,\, x_2$, while ${\mathbf x}$ corresponds to just $x_1$ and $x_2$.

In the representation we adopt for Dirac's algebra,  $\gamma_\mu$, $\mu = 0,
1, 2$, are Hermitian $2\times 2$ matrices, satisfying \mbox{$\{ \gamma_\mu
, \gamma_\nu \} \,=\, 2 \, \delta_{\mu\nu}$}.
Indices from the middle of the Greek alphabet, like $\mu$, $\nu$, \ldots ,
run from $0$ to $2$, while Roman ones can take the values $1$ and $2$.
The Dirac field has two spinorial components, and it can be used as a
building block for higher, reducible representations (this is indeed the
usual situation in graphene models). 

Regarding the specific form of the mass $M({\mathbf x})$, we shall restrict
ourselves in this work to configurations such that \mbox{$|M({\mathbf x})|
= m = {\rm constant}$} almost everywhere, changing sign along a
two-dimensional spacetime region ${\mathcal U}$ which, for static domain
walls, has the form 
\begin{equation}
{\mathcal U} \;=\; {\mathcal C} \times {\mathbb R}
\end{equation}
where ${\mathcal C}$ is a one dimensional region contained on ${\mathbb
R}^2$,  the $x_1, x_2$ plane.    
For static configurations, ${\mathcal C}$ is assumed, in this paper, to
correspond to  either a single curve, or to two disjoint regular curves. In
the next sections, we discuss those two cases separately.  

We conclude this Section by introducing the effective action,
$\Gamma(M)$:
\begin{equation}
	e^{- \Gamma(M)} \;=\; {\mathcal Z}(M) \;,
\end{equation}
where
\begin{equation}
	{\mathcal Z}(M) \;=\; \int {\mathcal D}\psi 
{\mathcal D}\bar\psi \; e^{-{\mathcal S}(\bar\psi,\psi,M)}  \;,
\end{equation}
is the Euclidean vacuum to vacuum transition amplitude.  From $\Gamma(M)$
one can obtain the vacuum energy $E$:
\begin{equation}\label{eq:defe0}
	E \;=\; \lim_{T \to \infty} \frac{\Gamma(M)}{T} \;, 
\end{equation}
where $T$ denotes the extent of the time-like coordinate $x_0$ (regarded
temporarily as finite but tending to infinity to extract the vacuum energy).

We also note that ${\mathcal Z}(M)$ can be formally written in terms of a
fermionic determinant; indeed,
\begin{equation}
	{\mathcal Z}(M) \;=\; \det {\mathcal D} \;.
\end{equation}
Note that the effective action is real; indeed, since we work in the
Euclidean formalism, any imaginary part of $\Gamma(M)$, had it existed,
should have been parity violating (since the imaginary part of the
action is parity violating).  But no parity violating functional
(either local or nonlocal) can be constructed in terms of just a scalar
function and its derivatives. 
On the other hand, having in mind its application to graphene, the
reducible representations used in that context for the fermions are such
that the action is explicitly real). 

Thus, having the above discussion in mind, we can write:
\begin{align}
{\mathcal Z}(M) &=\; \det {\mathcal D} \;=\; \det {\mathcal D}^\dagger
\;=\; \big[ \det {\mathcal D}^\dagger \det {\mathcal D} \big]^{1/2} \nonumber\\
&=\;  \big(\det {\mathcal H}\big)^{1/2} \;=\; \big(\det
\breve{\mathcal H})^{1/2} \;,
\end{align}	
where	
\begin{equation}
{\mathcal H} \;\equiv\;  {\mathcal D}^\dagger {\mathcal D} \;,\;\;\;
\breve{\mathcal H} \;\equiv\;  {\mathcal D} {\mathcal D}^\dagger \;.
\end{equation}
Taking into account our assumptions about $M(x)$, we see that:
\begin{eqnarray}\label{eq:defh}
	{\mathcal H} &=& -\partial^2 + m^2 - \not \! \partial M(x)  \nonumber\\
	\breve{\mathcal H} &=& -\partial^2 + m^2 + \not \! \partial M(x) \;. 
\end{eqnarray}
Since results that can be expressed in terms of $\Gamma$ are independent of
the sign of $M$, one can work with either ${\mathcal H}$ or
$\breve{\mathcal H}$; in the remainder of this paper, we use the
former. Thus,
\begin{equation}\label{eq:gammadet}
	\Gamma(M) \;=\; - \,\frac{1}{2}\, \log \det {\mathcal H}
	\;=\; - \,\frac{1}{2}\, {\rm Tr} \log {\mathcal H}\;.
\end{equation}

\section{A single domain wall}\label{sec:sw}
We consider here a time-independent $M$, i.e., \mbox{$M = M({\mathbf x})$},
which has a domain wall defect along a single closed curve ${\mathcal C}$. In
other words, $M({\mathbf x})$ jumps from $-m$ to $+m$ when crossing
${\mathcal C}$ (physical observables, like the vacuum energy, are
independent of the sign of the jump). 

In order to gain some insight into the nature of the system, let us define
first the domain wall implicitly, in terms of a {\em smooth\/} function
$F({\mathbf x})$, assumed to vanish with a non-zero gradient on
${\mathcal C}$.   Then we set \mbox{$M({\mathbf x}) = m \,
\sigma(F({\mathbf x}))$} ($m \geq 0$), where $\sigma$ denotes the sign
function.  Thus, we see that the operator ${\mathcal H}$ introduced in
(\ref{eq:defh}) has the form:
\begin{equation}
{\mathcal H} \;=\;  {\mathcal H}_0 \,+\, {\mathcal H}_I\;,
\end{equation}
with
\begin{equation}\label{eq:defh0hi}
{\mathcal H}_0 \;=\; - \, \partial^2 \,+\, m^2 \;,\;\;\;
{\mathcal H}_I \;=\; -  2 \, m \, \delta[F({\mathbf x})] \,\gamma_j
\partial_jF({\mathbf x}) \;,
\end{equation}
where $\delta$ denotes Dirac's delta function, and ${\mathcal H}_0$ is
proportional to the (omitted) $2 \times 2$ unit matrix.

By an application of the coarea formula, we note that ${\mathcal H}_I$ can
also be rendered in the form: \mbox{${\mathcal H}_I({\mathbf x}) = \gamma_i
	{\mathcal A}_i({\mathbf x})$}, where
\begin{equation}
	{\mathcal A}_i({\mathbf x}) \;=\; -  2 \, m \, \int d\tau \, 
	\big|\frac{d{\mathbf y}(\tau)}{d\tau}\big| \,
	\delta[{\mathbf x} - {\mathbf y}(\tau)] \;
	\hat{n}_i(\tau)	\;,
\end{equation}
with $\tau \to {\mathbf y}(\tau)$ a parametrization of ${\mathcal C}$, and
$\hat{n}_i(\tau)$ is the $i^{th}$ component of $\mathbf{\hat{n}}(\tau)$,
the unit normal~\footnote{Results for $\Gamma$ are independent of the
choice (inwards or outwards) for the normal.} to ${\mathcal C}$ at the
point ${\mathbf y}(\tau)$. Thus, 
\begin{align}
{\mathcal A}_i({\mathbf x}) &=\;  \epsilon_{ij} \; \chi_j({\mathbf x}) \nonumber\\
	\chi_i({\mathbf x}) &=\; - 2 m \, \int d\tau \; \delta[{\mathbf x}
	- {\mathbf y}(\tau)] \; \frac{dy_i(\tau)}{d\tau}\;.
\end{align}

\subsection{Effective action and self-energy}
As it should have been expected, it is far from trivial to calculate
the effective action, and therefore the self-energy, exactly for an
arbitrary closed curve ${\mathcal C}$.  Besides, one should expect the
existence of infinities, due to the assumption that the defects have zero
width. Those infinities can nevertheless be regulated by the introduction
of an UV cutoff which, in coordinate space, may be regarded as a non
vanishing width for the defect. In the next Section, when dealing with the
interaction between two defects, no cutoff dependence is expected in the
interaction energy, since this object is independent of the respective
self-energies.

One can attempt to implement different approximation schemes in order to
calculate $\Gamma$; the choice is determined, in the end, by the nature
of the configuration being studied.  The case we
consider here corresponds to small-amplitude deviations with respect to a
linear defect. More precisely, we assume that $F({\mathbf x}) = x_2 -
\varphi(x_1)$, expanding the effective action in powers of $\varphi$, which
is the deviation with respect to the $x_2 = 0$ straight line.
   
We first note that, with this choice of $F$:
\begin{equation}
	{\mathcal H}_I \;=\; - 2 m \, \delta( x_2 - \varphi(x_1) ) \, 
	\big( - \varphi'(x_1) \gamma_1 + \gamma_2 \big) \;,
\end{equation}
where a prime denotes derivative with respect to the argument.
Denoting now by $\Gamma_i$ the order-$i$ term in the expansion for
$\Gamma$:
\begin{equation}
\Gamma \;=\; \Gamma_0 \,+\, \Gamma_1 \,+\, \Gamma_2 \,+\, \ldots
\end{equation}
we also have the corresponding expansion for the energy:
\begin{equation}
E = E_0 \,+\, E_1\, + E_2 + \ldots \;,
\end{equation}
where $E_0$ amounts to an uninteresting infinite constant independent of
$\varphi$, which can be interpreted as coming from a linear energy density.
The divergence is present already at the level of the energy density, which
is cutoff dependent. 

Regarding the first and second order terms, we see that, as a consequence
of assuming that the functional expansion exists, they can be written as
follows:
\begin{eqnarray}
E_1 &=& \int dx_1 \, {\mathcal E}^{(1)}(x_1) \, \varphi(x_1)
\nonumber\\
E_2 &=& \frac{1}{2} \, \int dx_1 \int dy_1 \; {\mathcal
E}^{(2)}(x_1,y_1) \, \varphi(x_1) \, \varphi(y_1) \;.
\end{eqnarray}
Since the functional expansion coefficients ${\mathcal E}^{(1)}$ and
${\mathcal E}^{(2)}$ are independent of $\varphi$, they must be
translation invariant objects. Thus, ${\mathcal E}^{(1)} = {\rm constant}$
and ${\mathcal E}^{(2)}(x_1,y_1) = {\mathcal E}^{(2)}(x_1-y_1)$.

The first-order term then has the form:
\begin{equation}
	E_1\;=\; {\mathcal E}^{(1)}\; \int dx_1 \varphi(x_1) \;;
\end{equation}
namely, it depends only on the average value of the deformation $\varphi$.
Since this average value could be changed just by performing a rigid
translation of the defect along the $x_2$ direction, and the energy cannot
change under such a shift, we conclude that ${\mathcal E}^{(1)}$, and
therefore $E_1$, vanishes. We have checked this explicitly, by evaluating
$\Gamma_1$, which, recalling (\ref{eq:gammadet}), is given by:
\begin{equation}\label{eq:order1}
	\Gamma_1\;=\; - \frac{1}{2}\, 
	{\rm Tr}\Big[ \big({\mathcal H}_0\big)^{-1} \, {\mathcal H}_1
	\Big] \;.
\end{equation}
Here, ${\mathcal H}_l$ consistently denotes de order-$l$ term in an
expansion of ${\mathcal H}_l$. The ones appearing here are:
\begin{eqnarray}
{\mathcal H}_0 &=&- \partial^2 \,+\, m^2 \,-\, 2 m \gamma_2 \delta(x_2) \nonumber\\
{\mathcal H}_1 &=& 2 m \big[ \gamma_1 \delta(x_2) \,\varphi'(x_1) +
\gamma_2 \delta'(x_2) \,\varphi(x_1)  \big] \;.
\end{eqnarray}

The inverse of ${\mathcal H_0}$ is given by 
\begin{eqnarray}
	({\mathcal H_0})^{-1}(x,y) &=& \int \frac{d^2k_\parallel}{(2\pi)^2}
	\, e^{ i k_\parallel \cdot (x_\parallel - y_\parallel)} \,
	G(k_\parallel; x_2,y_2) \;, \nonumber\\
	G(k_\parallel; x_2,y_2) &=& G^+(k_\parallel; x_2,y_2) \, {\mathcal
P}^+ \,+\, G^-(k_\parallel; x_2,y_2) \, {\mathcal P}^- \;, \nonumber\\
G^\pm(k_\parallel; x_2,y_2) &=& \frac{1}{2 \Omega(k_\parallel)} 
\Big[ e^{- \Omega(k_\parallel) |x_2-y_2|} \pm
\frac{m}{\Omega(k_\parallel) \mp m}  e^{- \Omega(k_\parallel)
(|x_2|+|y_2|)}\Big]
\end{eqnarray}
with ${\mathcal P}^\pm = \frac{1 \pm \gamma_2}{2}$,  $v_\parallel
\equiv (v_0,v_1)$ and $\Omega(k_\parallel)\,=\,\sqrt{k_\parallel^2 + m^2}$.

Then, after some algebra, we find:
\begin{equation}
	{\mathcal E}^{(1)}\,=\, - m \, \int dx_2
	\int\frac{d^2k_\parallel}{(2\pi)^2}  \big\{[ G^+(k_\parallel;
	x_2,y_2) - G^-(k_\parallel; x_2,y_2)] \delta'(y_2)\big\}\Big|_{y_2
	\to x_2}\;=\; 0 \;,
\end{equation}
as expected.

We conclude this Section by dealing with the second order term; it is made
up of two contributions:
\begin{eqnarray}
	\Gamma_2 &=& \Gamma_{2,1} \,+\, \Gamma_{2,2} \nonumber\\
	\Gamma_{2,1} &=& - \frac{1}{2}\, {\rm Tr}\Big[ \big({\mathcal
	H}_0\big)^{-1} \, {\mathcal H}_2 \Big] \nonumber\\
	\Gamma_{2,2} &=&  \frac{1}{4}\, 
	{\rm Tr}\Big[ \big({\mathcal H}_0\big)^{-1} \, {\mathcal H}_1 
	\big({\mathcal H}_0\big)^{-1} \, {\mathcal H}_1 \Big]
\end{eqnarray}

The first contribution involves ${\mathcal H}_2$:
\begin{equation}
	{\mathcal H}_2 \;=\; - m \,\left[ \gamma_1 \delta'(x_2)
	\partial_1(\varphi(x_1))^2 + \gamma_2 \delta''(x_2)
(\varphi(x_1))^2 \right]  \;,
\end{equation}
and yields an energy with the form:
\begin{equation}
E_{2,1} \;=\; {\mathcal E}^{(2,1)} \int dx_1 (\varphi(x_1))^2
\end{equation}
where
\begin{equation}
{\mathcal E}^{(2,1)} \;=\; \frac{m}{2} \, \int dx_2
	\int\frac{d^2k_\parallel}{(2\pi)^2} \,\big\{[ G^+(k_\parallel;
x_2,y_2) - G^-(k_\parallel; x_2,y_2)] \delta''(y_2)\big\}\Big|_{y_2 \to
x_2} \;.
\end{equation}
An explicit evaluation shows that the object above is quadratically
divergent in the ultraviolet, namely, introducing an UV cutoff $\Lambda$,
\begin{equation}
	{\mathcal E}^{(2,1)} \;=\; c \, \frac{(m\Lambda)^2}{2} \;,
\end{equation}
with $c$ a dimensionless constant, which depends upon the regularization
approach. 
Thus, we conclude that the role of this term amounts to introducing 
a mass density proportional to $(m\Lambda)^2$  for the collective
degree of freedom.

The remaining term, $E_{2,2}$, can be evaluated and represented in Fourier
space, the result being a nonlocal quadratic functional in $\varphi$:
\begin{equation}
E_{2,2} \;=\; \frac{1}{2} \, \int \frac{dk_1}{2\pi} \,
{\widetilde{\mathcal E}}^{(2,2)}(k_1) \, |\widetilde{\varphi}(k_1)|^2 \;.
\end{equation}

From that expression, we can extract its local piece, quadratic in derivatives, which is
logarithmically divergent:
\begin{equation}
	E_{2,2}\;=\; m^2 \, \int \frac{d^2k_\parallel}{k_\parallel^2}
\, \int dx_1 (\varphi'(x_1))^2 \;, 	
\end{equation}
which may be thought of as generating  
a `tension'  for the domain wall.
Note that the existence of an infrared divergence in the momentum integral
can only proceed from the existence of a massless field, which we can
readily identify here as corresponding to the one predicted by the Callan
and Harvey mechanism.

\subsection{Coupling to an external gauge field}
When coupling the Dirac field to an external Abelian gauge field $A_\mu$,
we have to perform the following change in the operator ${\mathcal D}$:
\begin{equation}
	{\mathcal D} \;\to\; \not \!  \partial \,+\, i  e \not \!\! A(x) \,+\,
	M({\mathbf x}) \;.
\end{equation}
Then, assuming a rectilinear defect, the term of second order in $A_\mu$,
$\Gamma^{(2)}(A)$, will have the structure:
\begin{equation}
	\Gamma^{(2)}(A) \;=\; \frac{1}{2} \, \int d^3x \int d^3y \, A_\mu(x) 
	\Pi_{\mu\nu}(x,y) \, A_\nu(y) \;.
\end{equation}

It is convenient to perform a Fourier transform of the objects above with
respect to the $x_0$ and $x_1$ coordinates (the defect breaks translation
invariance along the $x_2$ axis):
\begin{equation}
	\Gamma^{(2)}(A) \;=\; \frac{1}{2} \, \int
	\frac{d^2k_\parallel}{(2\pi)^2} \, \int dx_2 \int dy_2 \, 
	\widetilde{A}_\mu(-k_\parallel; x_2) 
	\widetilde{\Pi}_{\mu\nu}(k_\parallel; x_2,y_2) \, \widetilde{A}_\nu(k_\parallel; y_2) \;
\end{equation}
A straightforward calculation shows that the vacuum polarization tensor
$\widetilde{\Pi}_{\mu\nu}$ is given by:
\begin{equation}
\widetilde{\Pi}_{\mu\nu}(k_\parallel; x_2,y_2)\;=\; - \,e^2\, 
\int \frac{d^2p_\parallel}{(2\pi)^2} \,
{\rm tr} \Big[ 
S(p_\parallel; y_2,x_2) \gamma_\mu 
S(p_\parallel + k_\parallel; x_2,y_2) \gamma_\nu
\Big] \;,
\end{equation}
with:
\begin{equation}
	S(p_\parallel;x_2,y_2) \;=\; 
	\Big( 
	-\gamma_2 \partial_{x_2} - i \not \! p_\parallel + m \sigma(x_2) 
	\Big)G(p_\parallel; x_2,y_2) \;.
\end{equation}
Having in mind to study the response of the system to an electric field along the
direction and location of the defect, we consider the components
$\widetilde{\Pi}_{\alpha\beta}$, with $\alpha,\,\beta$ in the $0, 1$ range,
and set $x_2=y_2=0$. After a rather lengthy calculation, we see that the
only surviving contributions to the vacuum polarization lead to:
\begin{align}
	& \widetilde{\Pi}_{\mu\nu}(k_\parallel; x_2,x_2) \;=\;  \frac{e^2}{4} \;
	\int \frac{d^2p_\parallel}{(2\pi)^2} \,
\big[ p_\alpha (p+k)_\beta + p_\beta (p+k)_\alpha - \delta_{\alpha\beta}
	p_\parallel \cdot (p_\parallel + k_\parallel) \big] \nonumber\\
	&\times 
	\Big[
		\frac{1}{(\Omega(p_\parallel)  - m)(\Omega(p_\parallel +
		k_\parallel)  - m)} \,+\,
		\frac{1}{(\Omega(p_\parallel)  + m)(\Omega(p_\parallel +
		k_\parallel)  + m)}\Big] \;.
\end{align}
It can be seen that the leading contribution proceeds from the first term
on the second line above. In particular, for large $m$:
\begin{equation}
\widetilde{\Pi}_{\mu\nu}(k_\parallel; x_2,x_2) \;\sim\;  
\big(\frac{e m}{2}\big)^2 \;
\int \frac{d^2p_\parallel}{(2\pi)^2} \,
\frac{p_\alpha (p+k)_\beta + p_\beta (p+k)_\alpha - \delta_{\alpha\beta}
p_\parallel \cdot (p_\parallel + k_\parallel)}{ p_\parallel^2  (p_\parallel +
		k_\parallel)^2} \;,
\end{equation}
which is the expression for the vacuum polarization in $1+1$ dimensions,
due to a massless fermion field. Thus we have verified, in a concrete
example, the presence of that mode in our treatment of the problem. 

\section{Two domain walls}\label{sec:dw}
In this section, the mass is assumed to have a purely spatial dependence,
with two domain-wall like defects, i.e., zero-width regions of space where the mass
passes through zero. One of those regions will be assumed to correspond to
a straight line, hereafter denoted by $L$, defined by $x_2=0$. 
The other defect, $R$, is assumed to correspond to a 
curve which represents a small departure from a line which is parallel to
$L$. We assume that it can be defined in terms of a single function
$\varphi(x_1)$, which specifies the distance, along $x_2$, of each point in $R$ to $L$.
Thus,
\begin{equation}
M({\mathbf x}) \;=\; m \; \sigma (x_2) \; \sigma(x_2 - \varphi(x_1))
\end{equation}
where, as before,  $\sigma$ denotes the sign function. $m$ is a positive constant
which defines the constant value of the absolute value of $M({\mathbf x})$,
as well as half the height of the jump in the mass at each defect.

The assumption about $R$ being a small departure from a straight line
parallel to $L$ is made more precise by assuming that the above introduced
function $\varphi(x_1)$ can be written as \mbox{$\varphi(x_1) = a +
\eta(x_1)$}, with $a>0$ and \mbox{$|\eta(x_1)|<< a$}.

The effective action $\Gamma(M)$, can then be expanded in powers of $\eta$, 
\begin{equation}\label{Gammaexp}
	\Gamma \;=\; \Gamma_0 \,+\,\Gamma_1 \,+\,\Gamma_2 \,+\ldots
\end{equation}
where the index denotes the order in $\eta$ of the corresponding term.
We will evaluate here the zeroth order, and describe the calculation of 
the first and second orders in the Appendix.

The zeroth order corresponds to setting $\eta=0$, so that the walls $L$ and $R$
will be located at $x_2=0$ and $x_2 = a$, respectively. The system has then
translation invariance along $x_1$, as well as time independence. The
effective action to this order will then diverge, since it will be
proportional to the extent of the time interval, $T$, and to $L_1$, the
length of the system along $x_1$, which should tend to infinity.
As usual, one can take care of that divergence by considering the effective
action per unit time and per unit length, a quantity which we shall denote
by ${\mathcal E}_0(a)$, and has the dimensions of an energy per unit
length. That quantity,  a function of $a$ and $m$, contains the
information about the interaction energy between the two domain walls, in
particular on the part of that function which does depend on $a$. 
Self-energy contributions are $a$-independent and will be discarded.  In other
words, since the force per unit length between $L$ and $R$ is proportional
to (minus) the derivative of $\Gamma_0$ with respect to $a$, we only keep
the terms which contribute to that observable.

$\Gamma_0(a)$ is formally given by a functional determinant: 
\begin{equation}
e^{-\Gamma_0(a)}\;=\; \det \big[\not \!  \partial \,+\, m(x_2) \big] 
\,=\, \det \big[-\not \!  \partial \,+\, M_0(x_2) \big] \;,
\end{equation}
where the second equality is a consequence of the reality of the energy.

Then,
\begin{equation}
e^{-\Gamma_0(a)} \;=\; \Big\{ \det\Big[ \big(- \not \! \partial
\,+\, M_0(x_2) \big) \big(\not \!
\partial \,+\, M_0(x_2) \big) \Big] \Big\}^{\frac{1}{2}}\;.
\end{equation}
Then we see, by Fourier transforming the dependence on the $x_0$ and $x_1$
coordinates, that:
\begin{equation}
	{\mathcal E}_0(a)\;=\; \lim_{T,L_1\to\infty}
\,\frac{\Gamma_0}{T L_1}\;=\;
	-\, \frac{1}{2} \,\int \frac{d^2k_\parallel}{(2\pi)^2} \log
\det {\mathcal K}
\end{equation}
where $k_\parallel \equiv (k_0, k_1)$, and ${\mathcal K}$ denotes a
functional matrix operator acting on functions of $x_2$:
\begin{eqnarray} 
{\mathcal K} &=& - \partial_2^2 + k_\parallel^2  + m^2 - \, \gamma_2 \,
\partial_ 2 M_0(x_2) \nonumber\\
 &=& - \partial_2^2 + \Omega^2(k_\parallel) +  2 m \, \gamma_2 \, [ \delta(x_2) 
 - \delta(x_2 - a) ] \,.
\end{eqnarray}

As expected, the problem has been reduced to the calculation of a reduced
fermionic determinant involving a non trivial dependence on $x_2$ only.
Besides, the $2 \times 2$ matrix structure can be straightforwardly  dealt with,
decomposing the problem into two scalar ones:
\begin{align}
{\mathcal E}_0(a) \;= \;  -\, \frac{1}{2} \,\int \frac{d^2k_\parallel}{(2\pi)^2}\Big\{ 
& \log \big(-\partial_2^2 + \Omega^2(k_\parallel) +  2 m  [
\delta(x_2) - \delta(x_2 - a) ] \big) \nonumber\\
+ & \log \big(-\partial_2^2 + \Omega^2(k_\parallel) -  2 m [
\delta(x_2) - \delta(x_2 - a) ] \big) \Big\} \;.
\end{align}
Regarding these two scalar problems, since they involve operators acting
non trivially only on one coordinate, they can be evaluated using
Gelfand-Yaglom theorem, in an identical fashion to the one presented
in~\cite{CcapaTtira:2011ga}.
Following the method applied in that reference, we get for each scalar
problem, the same contribution (each one independent of the sign of $m$).
The expression for the energy density thus becomes:
\begin{equation}\label{interplane}
{\mathcal E}_0(a) \;=\; - \,\int \frac{d^2k}{(2\pi)^2}\, 
\log \big[ 1  + \frac{m^2}{k_\parallel^2} e^{ - 2  \Omega(k_\parallel) a}
\big] \;.
\end{equation}
By a rescaling of the integration variables, we can write:
\begin{equation}
{\mathcal E}_0(a) \;=\; - \frac{g( m a)}{a^2} \;,
\end{equation}
\begin{figure}[h!]
\begin{center}
\includegraphics[width=10cm]{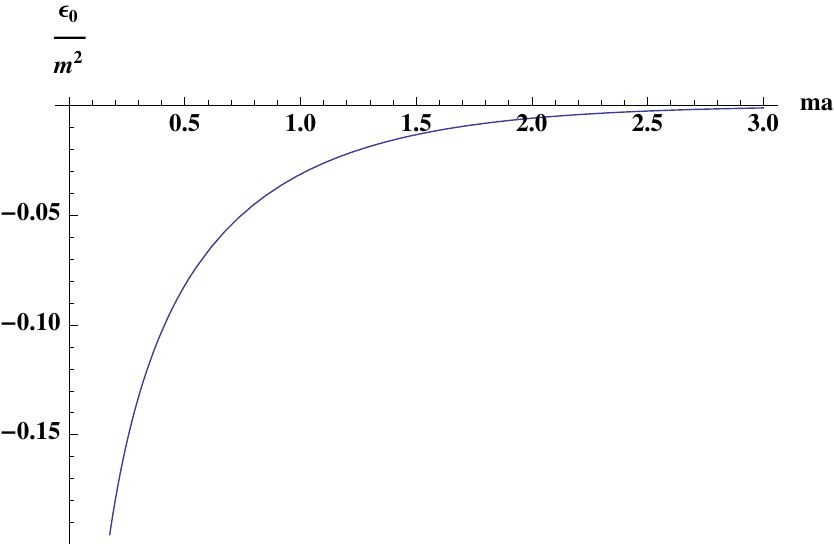}
\caption{Dimensionless energy density ${\mathcal E}_0/m^2$ as a function of the dimensionless distance $ma$ between planar domain walls. }
\label{fig1}
\end{center}
\end{figure}
where $g(x)$ is a dimensionless function of the only dimensionless function
that can be formed with $m$ and $a$. Its explicit form cannot be given in
a closed form, but nevertheless it can be written as an integral:
\begin{equation}\label{g}
g(x) \;=\; \frac{1}{4 \pi} \, \,\int_0^\infty \, du \; 
\log \big[ 1  + \frac{x^2}{u} e^{ - 2 \sqrt{ u + x^2}} \big] \;. 
\end{equation}
In Figure 1 we present a plot of the energy per unit area  ${\mathcal E}_0$ as a function 
of the distance between domain walls.  It is a monotonous function, and produces a force which is always attractive.
We have checked numerically that ${\mathcal E}_0/m^2$ diverges as $\log(ma)$ at short distances ($ma\ll 1$), and 
vanishes exponentially as $ma \exp(-2ma)$
in the opposite limit $ma\gg 1$. The behavior at large distances is typical for the vacuum interaction energy
associated to massive fields. It can also be obtained analytically by approximating $\Omega(k_\parallel) \approx m$ in Eq.(\ref{interplane}),
performing the integral up to a maximum value of $k_\parallel$ of order $m$, and then expanding the result
for $ma\gg 1$.

It is interesting to remark that the structure of the result for the vacuum interaction energy between domain walls,
Eq.(\ref{interplane}), is similar to those obtained for the Casimir effect for massive fermions between planar boundaries \cite{Elizalde}
or in the presence of $\delta$-potentials 
\cite{Losada}.
%

\section{Conclusions}\label{sec:conc}

We have computed the effects of quantum fluctuations of a
Dirac scalar field in $2+1$ dimensions on domain wall defects. For a single
defect, the vacuum energy is highly divergent. We can understand the origin
of the divergences as follows. Had we considered  a  theory in
which the fermion field is coupled to a dynamical scalar field whose
classical part generates a  smooth domain wall,  the vacuum polarization of
the fermion field would produce a renormalization of the mass of the scalar
field as long as a finite correction to the mass of the domain wall. In the
present paper, there is an additional source of divergences, because we are assuming
a zero-width domain wall. For a nonplanar wall, we have found that the vacuum energy contains
divergences that are proportional to $\varphi^2$ and to $\varphi'^2$.
This indicates that in a dynamical model for the wall, there
would be a renormalization of the mass and of the tension of the defect.
We have seen that, part of this renormalization appears to be due to the
fluctuations of the fermionic zero mode. We have verified this in an
independent fashion, by computing the vacuum polarization tensor on the
domain wall, that the virtual effects due to this mode are, indeed,
present.

For two domain walls, we have shown that vacuum energy induces a
Casimir-like force between defects. For planar walls, the force could  be
computed using standard techniques based in the Gelfand-Yaglom theorem; the result shows that it
is always attractive.
At short distances,  it is inversely proportional to the distance, while
vanishes exponentially at large distances. The divergences that occur in
the vacuum energy for a single defect are not present in the interaction
energy, which is moreover unambiguously defined.

We remark that, in the graphene case one should multiply our result corresponding to the attractive force by the proper number of two-component fermions.

We have also obtained explicit expressions for the interaction energy
between a planar wall and a slightly deformed wall (see the Appendix).  
As for the usual
Casimir effect, in this case the energy is a nonlocal functional of the
deformation.

\section*{Appendix}

In this Appendix we compute the first and second order terms in the expansion of the effective action
given in Eq.(\ref{Gammaexp}).

\subsection*{ First order}
The calculation of the first order term does not give a
new result, but it can be used as a consistency check for the previous
calculation. Indeed, the contribution of first order in $\eta$ has the
form:
\begin{equation}
\Gamma_1 \;=\; - \, {\rm Tr} \Big[ \big( \not \! \partial + M(x_2) \big)^{-1}
\, M_1 \Big]
\end{equation}
where \mbox{$M_1({\mathbf x}) \,=\, - 2 \, m \, \delta(x_2 - a) \,
\eta(x_1)$}.  
By taking the functional trace, after some algebra we see that the first
order term in the energy per unit length is:
\begin{eqnarray}
{\mathcal E}_1 \,&=&\, \lim_{T, L \to \infty} \frac{\Gamma_1}{T L_1}\nonumber\\ \;&=&\; 2 \, m \, \eta_0 
\int \,\frac{d^2k_\parallel}{(2\pi)^2}\, 
 {\rm tr} \Big[ \langle x_2 | \big( \gamma_2 \partial_2 + i \not \!
k_\parallel + M_0(x_2) \big)^{-1} |y_2\rangle\Big]\Big|_{x_2=y_2=a}
\end{eqnarray}
where `${\rm tr}$' denotes the trace over Dirac indices.  On the other hand, $\eta_0 \equiv
\frac{1}{L_1} \int dx_1 \eta(x_1)$ is the mean value of $\eta$.

Since the previous expression depends on $\eta$ only through the constant
$\eta_0$, it is not sensible to the details of its local space dependence.
Therefore, it can be obtained from the zeroth order expression. Indeed, one
should have the relation:
\begin{equation}
{\mathcal E}_1 \;=\; {\mathcal E}_0(a + \eta_0) \,- \, {\mathcal E}_0(a)
\;+ \; {\mathcal O}(\eta_0^2) \;, 
\end{equation}
so that the first order term we are about to calculate should be compared
with the one obtained by evaluating the derivative of the zeroth order term
with respect to $a$ and multiplying by $\eta_0$.

One can show that:
\begin{equation}
\langle x_2 | \big( \gamma_2 \partial_2 + i \not \!
k_\parallel + M_0(x_2) \big)^{-1} |y_2\rangle \;=\; [-\gamma_2 \partial_{x_2} - i \not \!
k_\parallel + M_0 (x_2)] \; \langle x_2 | {\mathcal K}^{-1} | y_2 \rangle \;,
\end{equation}
where ${\mathcal K}$ is the operator introduced in the calculation of the
zeroth order term.

The inverse of the scalar operator above can be obtained by using standard
techniques, and the result obtained by inserting it into the expression for
the first order term is consistent with the relation obtained between it
and the derivative of the zeroth order term. 

\subsection*{Second order}
The second order term $\Gamma_2$ receives two different contributions:
\begin{equation}
	\Gamma_2 \;=\; \Gamma_2^a \,+\, \Gamma_2^b \;.
\end{equation}
where:
\begin{equation}
\Gamma_2^a \;=\; \frac{1}{2} \, {\rm Tr} \Big[ 
\big( \not \! \partial + M_0(x_2) \big)^{-1} \, M_1 
\big( \not \! \partial + M_0(x_2) \big)^{-1} \, M_1 \Big]
\end{equation}
and
\begin{equation}
\Gamma_2^b \;=\; - \, {\rm Tr} \Big[ \big( \not \! \partial + M_0(x_2) \big)^{-1}
\, M_2 \Big]
\end{equation}
It may be seen that $\Gamma_2^b$ can, like the first order term, be derived
from the knowledge of the zeroth order term. In other words, it is only
sensitive to the average value of $\eta$. Thus, we shall concentrate on
$\Gamma_2^a$, since it is the only one that contains new information to
this order.

We see that:
\begin{eqnarray}
	\Gamma_2^a = \frac{1}{2} (2 m )^2  
	\int_{x_\parallel\,y_\parallel}\, 
& {\rm tr}& \Big[ 
\langle x | \big( \not \! \partial + M_0(x_2) \big)^{-1}|y\rangle
\eta(y_1) \nonumber\\
&\times&
\langle y | \big( \not \! \partial + M_0(x_2) \big)^{-1}|x\rangle
\eta(x_1) \Big]
\Big|_{x_2=y_2=a} \;.
\end{eqnarray}
The system is now time-independent but translation invariance along $x_1$
is not necessarily preserved. Thus, $\Gamma_2^a$ will produce a
contribution to the energy (total, nor the linear density), $E_2$, which in
Fourier space can be written as follows:
\begin{equation}
E_2 \,=\, \frac{1}{2}  \; \int \,dk_1\, |\tilde{\eta}(k_1)|^2\; f_2(k_1) \;
\end{equation}
with
\begin{equation}
	f_2 (k) \;=\; 4 m^2 \, \int \frac{d^2p_\parallel}{(2\pi)^2} \,
 {\rm tr}\Big[ \widetilde{G}(p_\parallel;a,a) \widetilde{G}(p_\parallel +
k_\parallel;a,a) \Big] 
\end{equation}
where we have introduced:
\begin{equation}
\widetilde{G}(p_\parallel;x_2,y_2)\;=\; \int d^2x_\parallel \, 
e^{-i  p_\parallel\cdot  x_\parallel} \,
\langle x_\parallel, x_2 | \big( \not \! \partial + M_0(x_2)
\big)^{-1}|0_\parallel, y_2\rangle \;.
\end{equation}

We can obtain a more explicit expression for
$\widetilde{G}(p_\parallel;x_2,y_2)$, as follows:
\begin{align}
\widetilde{G}(p_\parallel;x_2,y_2) = 
\Big[& ( - \partial_{x_2} - i \not \! p_\parallel + M_0(x_2) )  {\mathcal
G}_+(p_\parallel; x_2,y_2) \, {\mathcal P}_+ \nonumber\\ 
 + & ( \partial_{x_2} - i \not \! p_\parallel + M_0(x_2) )  {\mathcal
G}_-(p_\parallel;x_2,y_2) \, {\mathcal P}_- \Big]
\end{align}
where
\begin{equation}
{\mathcal G}^{\pm}(p_\parallel;x_2,y_2) = 
\langle x_2 | \big[ - \partial_{x_2}^2 + p_\parallel^2 \mp (\delta(x_2) -
\delta(x_2-a) ) \big]^{-1} |y_2 \rangle \;,  
\end{equation}
and ${\mathcal P}_{\pm} \equiv \frac{1 \pm \gamma_2}{2}$.

A rather length but otherwise straightforward calculation shows that:
\begin{eqnarray}
\widetilde{G}(p_\parallel;x_2,y_2) &=& - \frac{i \not \! p_\parallel}{2
\Omega(p_\parallel)} 
+
\frac{m}{2 p_\parallel^2} \, 
\frac{1 - e^{-2 \Omega(p_\parallel) a}}{1 + \frac{m^2}{p_\parallel^2} \,
e^{-2 \Omega(p_\parallel) a} } \nonumber\\
&\times&
\Big[ -i \not \! p_\parallel
(\frac{m}{\Omega(p_\parallel)} - \gamma_2) + 
\Omega(p_\parallel) \frac{e^{- 2 \Omega a}}{1 - e^{- 2 \Omega a}} \Big] \;.
\end{eqnarray}

Evaluating the Dirac trace, we see that the kernel $f_2$ is given by:
\begin{eqnarray}\label{f2}
f_2(k_\parallel) &=& 2 m^2 \; \int \frac{d^2p_\parallel}{(2\pi)^2} \, \Big\{ 
p_\parallel \cdot (p_\parallel + k_\parallel) 
\big[\frac{m^2}{p_\parallel^2 (p_\parallel + k_\parallel)^2} \, 
\big( 1- \frac{m^2}{\Omega(p_\parallel) \Omega(p_\parallel +
k_\parallel)}\big) \nonumber\\
&\times&  \big( 1 - ( 1 + \frac{m^2}{p_\parallel^2}) B(p_\parallel)\big) 
\big( 1 - ( 1 + \frac{m^2}{(p_\parallel + k_\parallel)^2})
B(p_\parallel+k_\parallel)\big) \nonumber\\
& -& \frac{1}{\Omega(p_\parallel) \Omega(p_\parallel + k_\parallel)} \,
\Big( 1 + \frac{m^2}{p_\parallel^2} \big( 1 - ( 1 +
\frac{m^2}{p_\parallel^2}) B(p_\parallel)\big) \nonumber\\ 
&+& \frac{m^2}{(p_\parallel + k_\parallel)^2} \big( 1 - ( 1 + \frac{m^2}{(p_\parallel +
k_\parallel)^2})
B(p_\parallel+k_\parallel)\big) \Big) \Big] \nonumber\\
&+& \frac{m^2}{p_\parallel^2 (p_\parallel + k_\parallel)^2}
\Omega(p_\parallel) \Omega(p_\parallel + k_\parallel) B(p_\parallel)
B(p_\parallel+k_\parallel) \Big\} \;,
\end{eqnarray}
where we have introduced $B(p_\parallel)= \big(e^{ 2 \Omega(p_\parallel) a} +
\frac{m^2}{p_\parallel^2}\big)^{-1}$.
We have checked that this kernel is indeed finite, so that the expansion is, at least up to this order, well defined.

\section*{Acknowledgements}
This work was supported by ANPCyT, CONICET, and UNCuyo.


\newpage
\begin{thebibliography}{bib}
\bibitem{Ho:1984zz} 
T.~L.~Ho, J.~R.~Fulco, J.~R.~Schrieffer and F.~Wilczek,
Phys.\ Rev.\ Lett.\  {\bf 52}, 1524 (1984).\\
M.~Stone, A.~Garg and P.~Muzikar,
Phys.\ Rev.\ Lett.\  {\bf 55}, 2328 (1985).
\bibitem{Witten:1984eb} 
E.~Witten, Nucl.\ Phys.\ B {\bf 249}, 557 (1985).
\bibitem{Callan:1984sa} 
C.~G.~Callan, Jr. and J.~A.~Harvey,
Nucl.\ Phys.\ B {\bf 250}, 427 (1985).
\bibitem{Wen:1990se} 
X.~G.~Wen, Phys.\ Rev.\ B {\bf 41}, 12838 (1990).
\bibitem{BordagSG}
 M.~Bordag and J.~M.~Munoz-Castaneda,
  J.\ Phys.\ A {\bf 45}, 374012 (2012).
\bibitem{Kenneth}
 O.~Kenneth and I.~Klich,
  Phys.\ Rev.\ B {\bf 78}, 014103 (2008).
  \bibitem{Manton}
  N.~S.~Manton and P.~Sutcliffe,
 {\it Topological solitons}, Cambridge University Press (2004).  
\bibitem{Cortijo:2012}A.~Cortijo, F.~Guinea and M.~A.~H.~Vozmediano,
J.\ Phys.\ A {\bf 45}, 383001 (2012).
\bibitem{Ebert:2014}D.~Ebert1, V.~Ch.~Zhukovsky and E.~A.~Stepanov,
J.\ Phys.\: Condensed Matter {\bf 26}, 125502 (2014).
\bibitem{Semenoff}
  G.~W.~Semenoff, V.~Semenoff and F.~Zhou,
  Phys.\ Rev.\ Lett.\  {\bf 101}, no. 8, 087204 (2008).
\bibitem{CcapaTtira:2011ga} 
  C.~Ccapa Ttira, C.~D.~Fosco and F.~D.~Mazzitelli,
  J.\ Phys.\ A {\bf 44}, 465403 (2011)
  [arXiv:1107.2357 [hep-th]].
\bibitem{Elizalde} 
  E.~Elizalde, F.~C.~Santos and A.~C.~Tort,
  Int.\ J.\ Mod.\ Phys.\ A {\bf 18}, 1761 (2003).
  \bibitem{Losada}
  C.~D.~Fosco and E.~L.~Losada, Phys.\ Rev.\ D {\bf 78}, 025017 (2008).
  \end{thebibliography}
\end{document}